\documentclass[a4paper]{jpconf}
\usepackage{graphicx}
\usepackage{wrapfig}
\RequirePackage{lineno}

\newcommand{\pT}{$p_{T}$}

\newcommand{\sNN}{$\sqrt{s_{_{NN}}}$}

\newcommand{\DirPho}{$\gamma_{\rm dir}$}
\newcommand{\piZro}{$\pi^{0}$}

\newcommand{\ptAssoc}{$p_{T}^{\mathrm{assoc}}$}
\newcommand{\ptTrig}{$p_{T}^{\mathrm{trig}}$}
\newcommand{\zT}{$z_{T}$}

\newcommand{\IAApiZro}{$I_{AA}^{\pi^{0}}$}
\newcommand{\IAAg}{$I_{AA}^{\gamma_{dir}}$}

\begin{document}

\title{Direct photon-hadron correlation measurement and a way towards photon-triggered jets at RHIC}

\author{Nihar Ranjan Sahoo (For the STAR Collaboration)}

\address{Cyclotron Institute, Texas A$\&$M University, USA}

\ead{nihar@rcf.rhic.bnl.gov}

\begin{abstract}
 We report the results of  \DirPho- and \piZro--hadron azimuthal correlations as a measure of
  the away-side jet-like correlated yields in central Au+Au and p+p collisions at  \sNN~=200 GeV in the STAR experiment. The charged-hadron
per-trigger yields at  mid-rapidity with respect to high-\pT~ \DirPho~and \piZro~ 
in central Au+Au  collisions are compared with p+p collisions. Within uncertainties, the same \zT  (~\ptAssoc/\ptTrig) dependence of the suppression is observed for
\DirPho- and \piZro-  triggers. The results are compared with energy-loss model predictions. The $\gamma-\rm jet$ measurements can provide further understanding on the redistribution of in-medium energy loss. Ongoing $\gamma-\rm jet$ studies in the STAR experiment are also discussed.
\end{abstract}

\section{Introduction}
The azimuthal correlation of charged hadrons with respect to a direct-photon (\DirPho) trigger was proposed as a promising probe to study the mechanisms of parton energy loss~\cite{Wang_Huang_Sarcevic}. Since a \DirPho-trigger escapes without interacting with the medium, it approximates the initial energy of the recoil parton, which is subject to medium modifications. The recoil parton of a $\gamma_{\rm dir}$ trigger is a quark in leading-order QCD processes, whereas that of a high-\pT~$\pi^{0}$ trigger can be a quark or a gluon. In addition, a coincidence measurement of $\pi^{0}$ is biased to have been produced near the surface of the medium, while that of a $\gamma_{\rm dir}$ does not suffer from the same bias~\cite{Renk,ZOWW}. Hence the comparison between the suppression of per-trigger away-side associated yields of $\gamma_{\rm dir}$  to those of $\pi^{0}$ triggers should exhibit differences due to both the color-factor dependence and the path-length dependence of energy loss.

\vspace{-5pt}
 \section{STAR Detectors and Experimental techniques}
The data were taken by the Solenoidal Tracker at RHIC (STAR) experiment in 2011 and 2009 for Au+Au and $p+p$ collisions at \sNN~= 200 GeV, respectively. The Time Projection Chamber (TPC) is the main charged-particle tracking detector providing track information for the charged hadrons with $|\eta| < 1.0$~\cite{STAR_TPC}.  Events having a transverse energy in a BEMC ~\cite{STAR_BEMC} cluster $E_{T} >$ 8~GeV, with $|\eta|\leq 0.9$, are selected for this analysis. The associated charged particles are selected in range 1.2 GeV/c $<$ \ptAssoc, whereas \piZro~and \DirPho~are triggered within 12 $<$ \ptTrig~ $<$ 20 GeV/c. In order to distinguish a \piZro, which at high \pT~predominately decays to
two photons with a small opening angle, from a single-photon cluster, a transverse shower-shape
analysis is performed. A detailed discussion about the transverse shower profile ($\rm TSP$) method and experimental techniques used in this analysis can be found in Ref.~\cite{STAR_GammaHadron_PLB,STAR_GJet}.
\begin{figure}[htbp]
\begin{center}
  \includegraphics[width=0.4\textwidth]{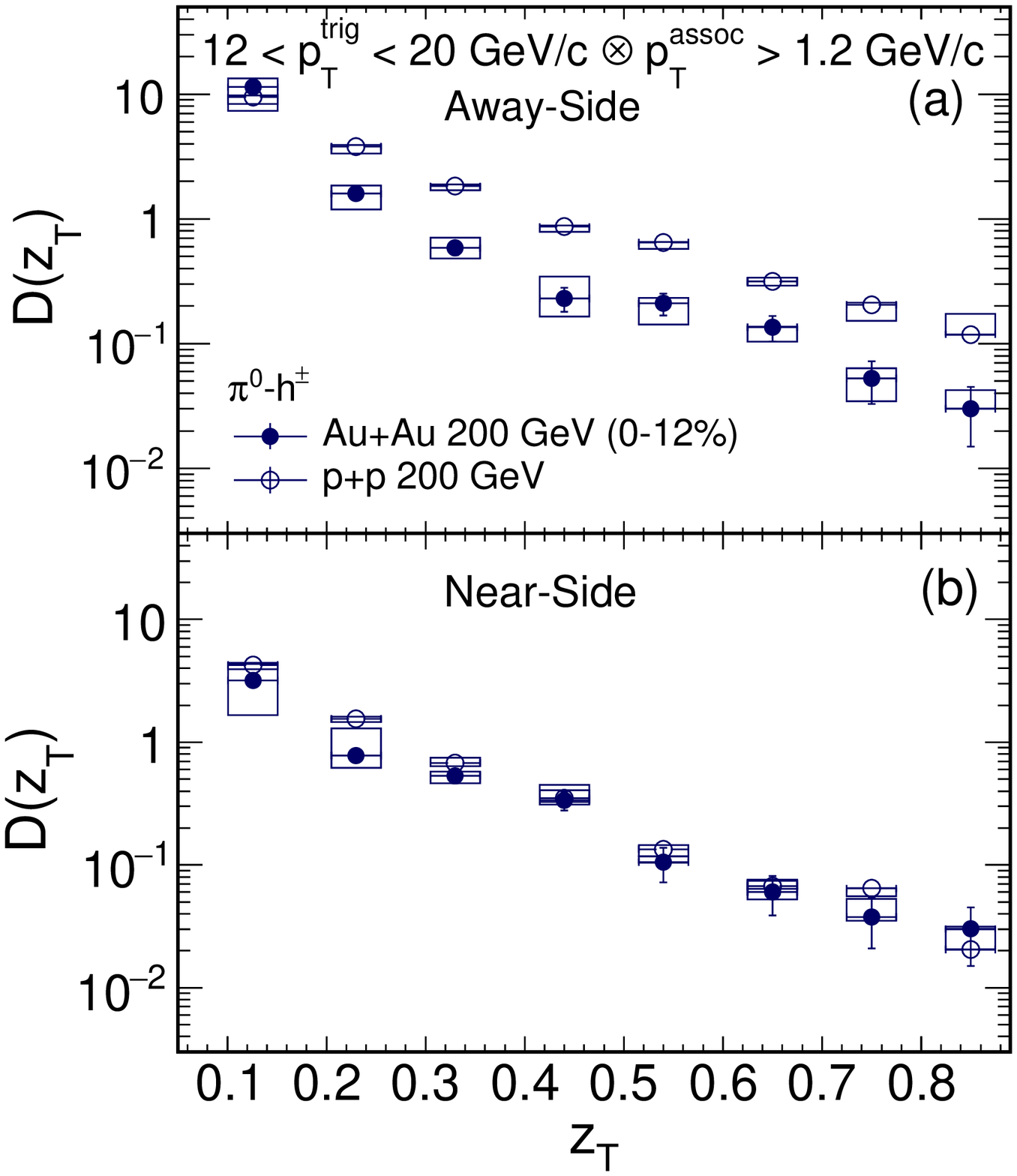}
   \includegraphics[width=0.4\textwidth]{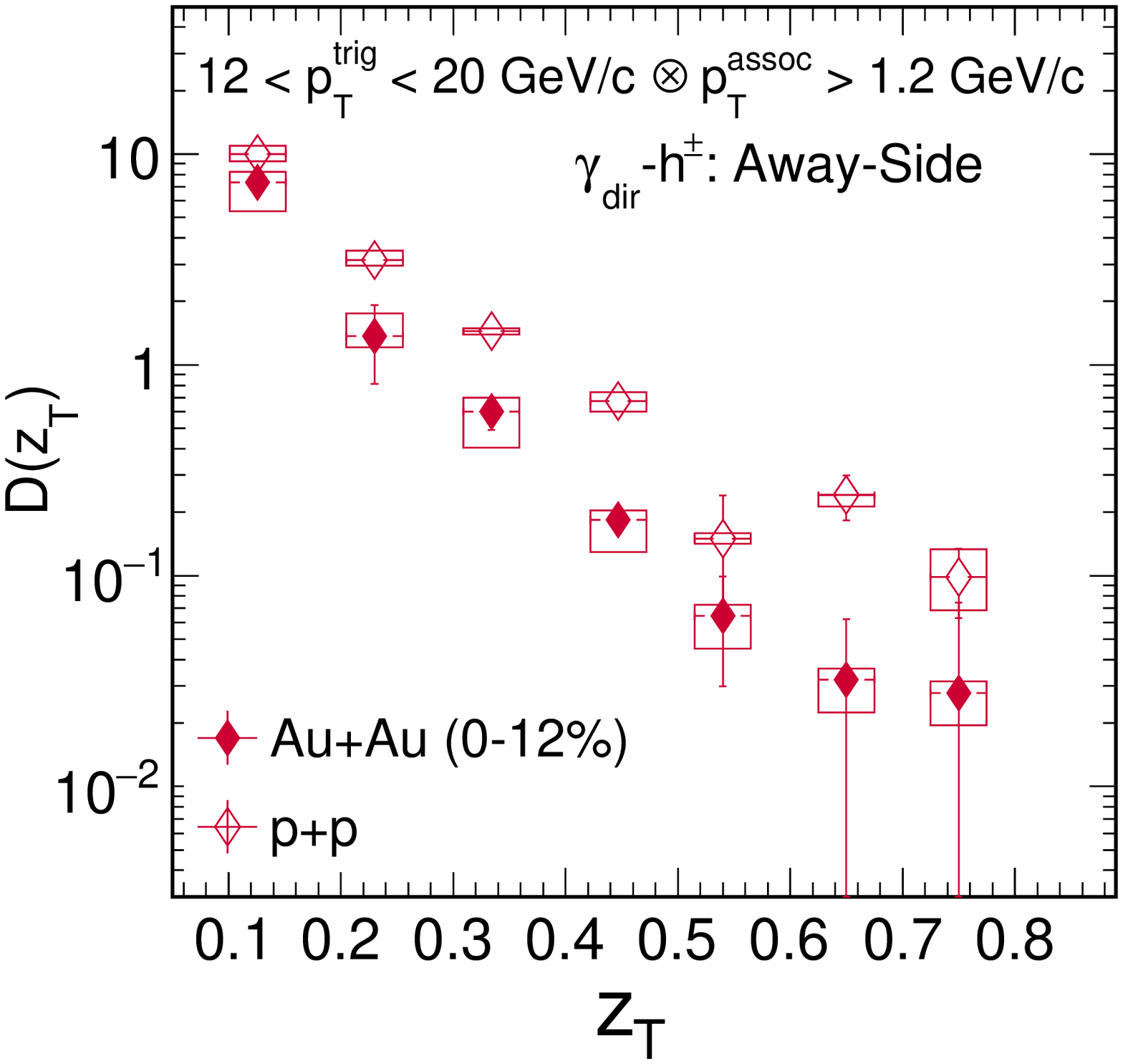}		
  \caption{(Color online.) Left panel: The \zT~dependence of \piZro-$h^{\pm}$
    away-side (a) and near-side (b) associated charged-hadron yields
    per trigger for Au+Au at 0-12$\%$ centrality
  (filled symbols) and  $p+p$ (open symbols) collisions at \sNN~= 200 GeV. Right panel: The \zT~dependence of 
  \DirPho-$h^{\pm}$ away-side associated charged-hadron yields per
  trigger for Au+Au at 0-12$\%$ centrality (filled diamonds) and
  $p+p$ (open diamonds) collisions~\cite{STAR_GammaHadron_PLB}. Vertical lines represent statistical
  errors, and the vertical extent of the boxes represents systematic uncertainties.}
\label{Fig1}
\vspace{-10pt}
\end{center}
\end{figure}
\begin{wrapfigure}{r}{0.5\textwidth}
\vspace{-10pt}
\includegraphics[width=0.5\textwidth]{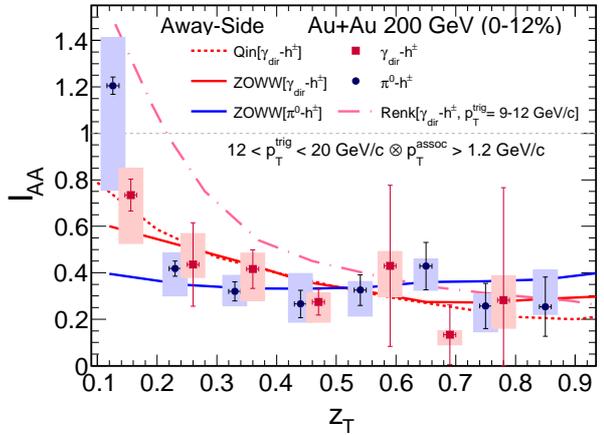}
\vspace{-6pt}
\caption{(Color online.) The  \IAAg~for \DirPho~(red  squares) and \IAApiZro~for \piZro~(blue circles) triggers are plotted as a function of \zT. The points for \IAAg~ are shifted by $+0.03$ in \zT~for visibility. The vertical line and shaded boxes represents statistical and systematic errors, respectively~\cite{STAR_GammaHadron_PLB}. The curves  represent theoretical model predictions~\cite{ZOWW,Qin,Wang,YAJEM}. }
\label{Fig2}
\end{wrapfigure}

\section{ Results: \DirPho- and \piZro--hadron azimuthal correlation}
The integrated away-side and near-side charged-hadron yields per \piZro~trigger, $D(z_{T})$, are plotted as a function of
\zT, both for Au+Au (0-12$\%$ centrality) and $p+p$ collisions, in the left panel of Fig.~\ref{Fig1}. The away-side $D(z_{T})$
for \DirPho~triggers as a function of \zT~for central Au+Au and minimum-bias $p+p$
collisions is shown in the right panel of Fig~\ref{Fig1}.  Yields of the away-side associated charged hadrons are suppressed, in Au+Au relative to $p+p$, at all \zT~except in the low \zT~region both for \DirPho~ and \piZro~trigger. On the other hand, no suppression is observed on the near-side in Au+Au, relative to $p+p$ collisions, due to the surface bias 
imposed by triggering on a high-$p_T$ \piZro. 
\begin{figure}[htbp]
\begin{center}
\includegraphics[width=0.6\textwidth]{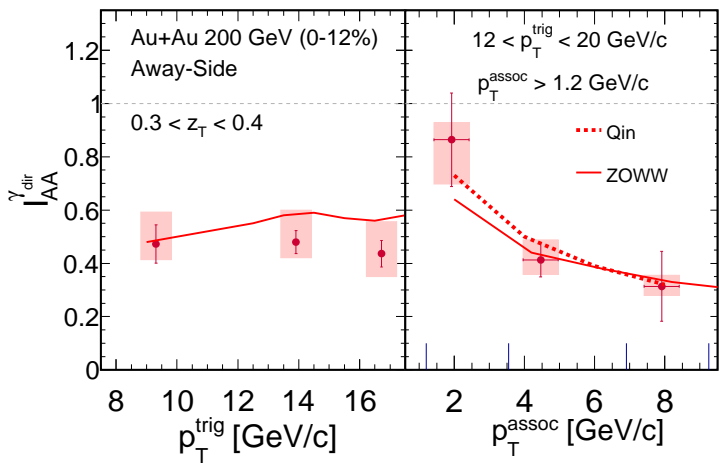}
\caption{(Color online.) The values of \IAAg~are plotted as a function of \ptTrig~(left panel) and \ptAssoc~(right panel)~\cite{STAR_GammaHadron_PLB}. The vertical line and shaded boxes represents statistical and systematic errors, respectively. The curves  represent theoretical model predictions~\cite{ZOWW,Qin,Wang}. }
\label{Fig3}
\end{center}
\end{figure}
  In order to quantify the medium modification for \DirPho- and \piZro-triggered recoil jet production as a function of \zT, the ratio defined as $ I_{AA} = \frac{D(z_{T})^{AuAu}}{D(z_{T})^{pp}},$ of the per-trigger conditional yields in Au+Au to those in $p+p$ collisions is calculated. 
  
\vspace{-5pt}
\begin{wrapfigure}{r}{0.5\textwidth}
\includegraphics[width=0.5\textwidth]{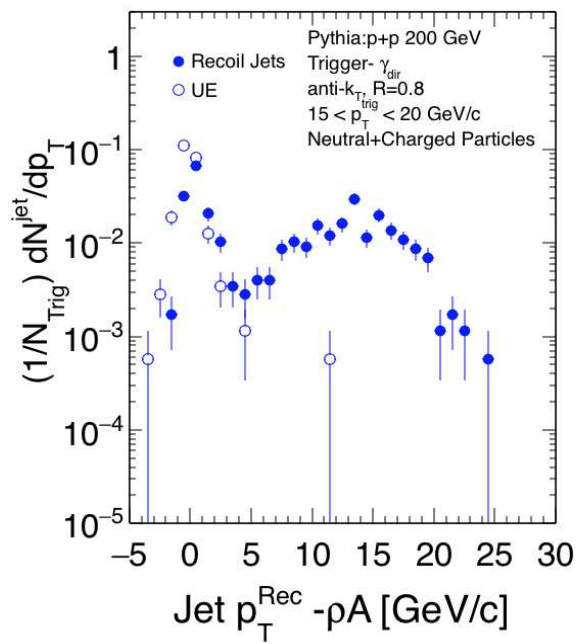}
\caption{(Color online.) \DirPho~triggered recoil (filled symbols) and uncorrelated (open symbols) reconstructed full jet transverse momentum spectra from PYTHIA simulation for p+p collisions at \sNN~= 200 GeV.  }
\vspace{-15pt}

\label{Fig4}
\end{wrapfigure}
 
  Figure~\ref{Fig2} shows the away-side medium modification factor for \piZro~triggers (\IAApiZro) and \DirPho~triggers (\IAAg), as a function of \zT. \IAApiZro~and \IAAg~show similar suppression within uncertainties. At low \zT~(0.1$<$\zT$<$0.2), both \IAApiZro~and \IAAg~show an indication of less suppression than at higher \zT.  
This observation is not significant in the \zT-dependence of $I_{AA}$ because the uncertainties in the lowest \zT~bin are large.
However, when $I_{AA}$ is plotted vs. \ptAssoc~(in Figure~\ref{Fig3}), the conclusion is supported with somewhat more significance. 
At high \zT, both \IAApiZro~and \IAAg~show a factor $\sim 3-5$ suppression. The ZOWW calculation also predicts \IAAg~as a function of \ptTrig~to be approximately flat in this range~\cite{ZOWW}. The YaJEM model predicts that at low \zT =0.2, \IAAg~= 1 and rises above unity even at lower \zT, although at lower triggered $p_{T}$ range 9-12 GeV/c. 

The values of \IAAg~are plotted as function of \ptAssoc in Fig.~\ref{Fig3}. It shows that the low-\ptAssoc~hadrons on
the away-side are not as suppressed as those at high \ptAssoc.  Both model predictions shown~\cite{ZOWW,Qin}, which do not include the redistribution of lost energy, are in agreement with the data. \IAAg~shows no sensitivity as a function of \ptTrig~, for 0.3$<$ \zT~$<$ 0.4, indicating that away-side parton energy loss is not sensitive to the initial energy of the parton in the range of 8-20 GeV/c.

\section{ Simulation study on $\gamma+\rm jet$} 
\vspace{4pt}
We have performed a feasibility study for 
$\gamma+\rm jet$ measurement in the kinematic acceptance for the STAR detector system using PYTHIA simulations. The $\gamma$-triggered events
are generated within 15 $<$ \ptTrig ~$<$ 20 GeV/c and all tracks are selected within 0.2 $< p_{\rm T}^{\rm track} <$ 20 GeV/c. Recoil full (including all charged and neutral particles) jets are reconstructed using the anti-$k_{T}$ algorithm~\cite{antikT,FASTJET} for a jet resolution parameter of $R =$0.8. Recoil jets are selected within $|\Delta \phi - \pi|  < \pi/4$, where $|\Delta \phi|$ is the difference between triggered \DirPho~ and reconstructed jet azimuth. The uncorrelated reconstructed full jet (UE) are selected within $ \Delta \phi \in [\pi/4, \pi/2]$ and $ \Delta \phi \in [ 3\pi/2,7\pi/4]$. The estimated background energy density ($\rho$) scaled by jet area ($A$) is subtracted from each reconstructed jet's raw transverse momentum, and the
corrected value, $p_{T,jet}^{Rec} - \rho A$, is shown in Fig.~\ref{Fig4}. Similar analyses with respect to hadronic trigger objects are discussed in Ref~\cite{ALICE,STAR_hjet}.   The triggered recoil jet peak, within 15 $<$ $p_{T,jet}^{Rec} - \rho A$ $<$ 20 GeV/c, can be seen around triggered \ptTrig~of $\gamma$. In the STAR experiment, similar measurements for charged and full jets reconstruction using TPC and BEMC detectors is underway within wider acceptance in $|\eta| < 1$ and $2\pi$-azimuth.

\section{ Summary and Outlook}

Within experimental uncertainty, both \IAApiZro~and \IAAg~show similar levels of suppression with the expected differences due to the color-factor
effect and the path-length dependence of in-medium energy loss not manifesting themselves. At high \zT (\ptAssoc), \IAAg~shows high suppression than at low \zT (\ptAssoc).
There is no trigger-energy dependence observed in the suppression of \DirPho-triggered yields, suggesting little dependence for energy loss on the initial parton energy, in the range of \ptTrig ~= 8-20 GeV/c. A semi-inclusive study of jets correlated with high-\pT~$\gamma$ is underway in the STAR experiment to explore more on the jet energy loss in the medium created at RHIC.


\end{document}